\documentclass[twocolumn,superscriptaddress]{revtex4-2}%
\usepackage{amsmath}%
\usepackage{amsfonts}%
\usepackage{amssymb}%

\usepackage{graphicx}

\usepackage{fullpage}

\usepackage{dsfont}
\usepackage{hyperref}
\usepackage{bm}
\usepackage{multirow}
\usepackage{color}
\usepackage{float}
\usepackage{hyperref}
\usepackage[colorinlistoftodos]{todonotes}

\newcommand{\ketv}[1]{\vert #1 \rangle}

\renewcommand{\>}{\rangle}

\begin{document}

\title{Electron dynamics in a 2D nanobubble: \\ A two-level system based on spatial density}

\author{Roberto Rosati}
\affiliation{Department of Physics, Philipps-Universit\"{a}t Marburg, Renthof 7, D-35032 Marburg, Germany}
\email{rosatir@uni-marburg.de}

\author{Frank Lengers}
\affiliation{Institut of Solid State Theory, University of M\"{u}nster, 48149 M\"{u}nster, Germany}

\author{Christian Carmesin}
\author{Matthias Florian}
\affiliation{Institute for Theoretical Physics, University of Bremen, P.O. Box 330440, 28334 Bremen, Germany}

\author{Tilmann Kuhn}
\affiliation{Institut of Solid State Theory, University of M\"{u}nster, 48149 M\"{u}nster, Germany}

\author{Frank Jahnke}
\author{Michael Lorke}
\affiliation{Institute for Theoretical Physics, University of Bremen, P.O. Box 330440, 28334 Bremen, Germany}

\author{Doris E. Reiter}
\affiliation{Institut of Solid State Theory, University of M\"{u}nster, 48149 M\"{u}nster, Germany}

\begin{abstract}
Nanobubbles formed in monolayers of transition metal dichalcogenides (TMDCs) on top of a substrate feature localized potentials, 
in which electrons can be captured.
We show that the captured electronic density can exhibit a non-trivial spatiotemporal dynamics,
 whose movements can be mapped to states in a two-level system illustrated as points of an electronic Poincar\'e sphere. These states can be fully controlled, i.e, initialized and switched, by multiple electronic wave packets. Our results could be the foundation for novel implementations of quantum circuits.
\end{abstract}

\maketitle

\section{Introduction}
Monolayers of transition metal dichalcogenides (TMDCs) are atomically thin two-dimensional (2D) semiconductors
 attractive for several applications in electronics and optoelectronics.\cite{Wang12,Mak16,Wang18,Mueller18} 
Due to their two-dimensional nature and the strong dependence of the material properties on strain and substrates, 
TMDC monolayers can host localized potentials of different extensions, ranging from few 
Angstroms for atomic defects \cite{Zhou13,Zhang17c,Klein19b,Chu20} to several hundreds of nanometers for 
strain-induced potentials.\cite{Tonndorf15,Kern16,Branny17,Palacios17,Rosenberger19} 
They are receiving increasing interest in the context of quantum information processing, because localized excitons may serve as single photon sources \cite{Tonndorf15}. Localized potentials can be either deterministically induced, 
for example by patterned substrates \cite{Kumar15,Kern16,Branny17,Palacios17,Rosenberger19} and 
helium irradiation,\cite{Klein19b}
or they build up naturally in the form of defects and disorder \cite{Zhou13,Zhang17c,Chu20} or nanobubbles.\cite{Carmesin19,Khestanova16,Shepard17,Chirolli19,Tyurnina19,Smiri21,Zhang20,Darlington20,Rodriguez21} Bubbles are naturally formed in experiments when depositing the 2D material 
on a substrate,\cite{Khestanova16,Shepard17,Tyurnina19,Zhang20,Darlington20,Rodriguez21,Blundo21} similar to the everyday process of air-bubbles forming when putting a plastic foil on glass. 
Bubbles in various monolayers and of different dimensions can also be obtained via bottom-up approaches such as ionic irradiations. \cite{Blundo21,Villarreal21}
Nanobubbles are able to induce an energy confinement due to the interplay of strain and local screening variations,\cite{Carmesin19}  resulting in a non-trivial  circular shape recently observed in experiments.\cite{Darlington20,Rodriguez21} Besides being extensively studied via optical response,\cite{Kumar15,Kern16,Branny17,Palacios17,Rosenberger19,Feierabend19,Sortino20,Thompson20} localized potentials can impact also the transport and associated recombination in different hosting 2D materials \cite{Leconte17,Zipfel20,Chu20} and even lead to excitonic funneling for larger potentials.\cite{Cordovilla18,Moon20,Harats20,Gelly21,Koo21}

\begin{figure}[t!]
\centering
 \includegraphics[width=1\columnwidth]{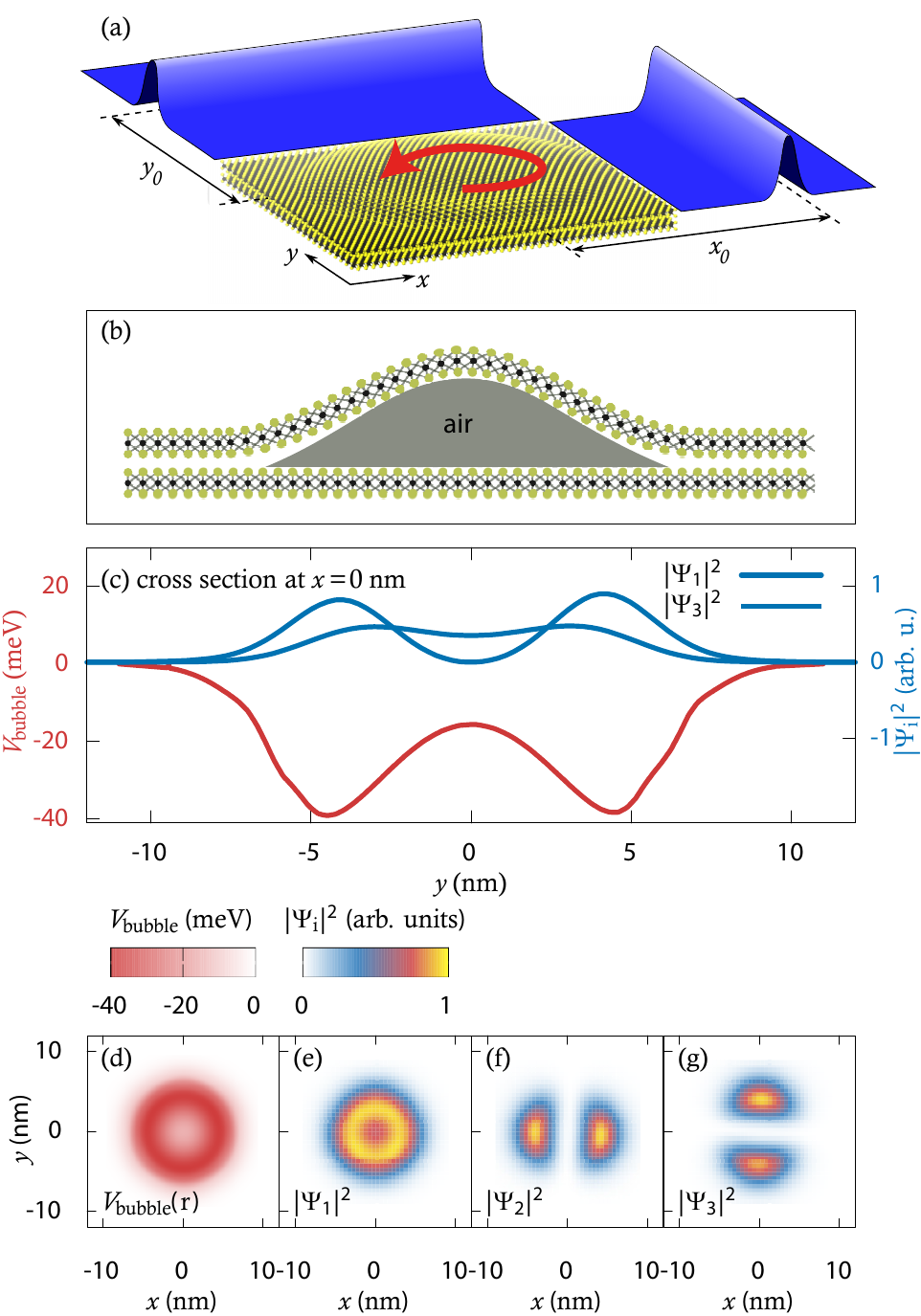}
\caption{(a) Set-up of our system: A nanobubble in a TMDC monolayer is impinged by wave packets from the $x$- and $y$-direction, respectively, which induces and controls the dynamics of the density in the region of the nanobubble. (b) Sketch of the microscopic setup of the nanobubble. (c) Cross section along the line $x=0$ of the system showing the microscopically calculated potential (red) and two resulting bound state wavefunctions (light and dark blue). (d) Color plot of the potential and (e)-(g) wave functions of the lowest three states.}
\label{fig1n}
\end{figure}

In localized structures a non-trivial spatiotemporal dynamics can be induced by capture from an electronic wave packet moving in 
the 2D monolayer.\cite{Rosati18} In particular the carrier capture can be mediated by the emission of phonons, which interact strongly with the electrons 
in the TMDC monolayer.\cite{Sohier16,Jin14}
The crucial property behind this peculiar dynamics is the locality of the carrier capture process, which was already 
studied in quantum wire-dot geometries.\cite{Reiter06,Reiter07,Rosati17} 
In this paper we combine two state-of-the art theoretical approaches to describe the dynamics of electronic wave packets in a realistic model of a 
TMDC hosting a nanobubble. 
We show that the resulting dynamics of the captured distribution can be well 
controlled by a combination of impinging wave packets, Fig.~\ref{fig1n}(a), allowing
movements ranging from linear along varying axes to circular. 
We show that all movements can be initialized, switched and mapped on a Poincar\'e sphere, analogous to the polarization states of light. This means that the electronic movement of the captured density 
inside the nanobubble can be understood as a two-level system, which could be used as quasi-stationary quantum state.

\section{System set-up}

A sketch of our system is shown in Fig.~\ref{fig1n}(a): %
In a TMDC monolayer a nanobubble (yellow-gray area) is formed. To calculate the potential created by this nanobubble, 
we utilize microscopic calculations, based on atomic force field and subsequent tight-binding simulations (see also Supporting Information). 
The method is based on the requirements of covering strain as well as local changes of the screening and electronic hybridization on the nanometer scale. 
To investigate the interplay between TMDC layer deformations and carrier confinement, we model a TMDC layer encapsulating an air bubble with atomic resolution 
using a million-atom supercell as shown in Fig.~\ref{fig1n}(b) similar to Ref.~\cite{Carmesin19}. Considering as a material a MoS$_2$ monolayer, 
the effects of the nanobubble result in the formation of a zero-dimensional (0D) confinement potential in the conduction band as 
plotted in Figs. \ref{fig1n}(c) and (d): It is
mostly radially
symmetric and shows the values of maximum depth (about -40 meV) distributed on a ring, rather than at the center.
This is the result of two different spatially-varying contributions, i.e.  dielectric constant and strain, which are stronger at the center and tail of the nanobubble, respectively.\cite{Carmesin19} This non-trivial shape has been recently experimentally observed in TMDC-based structures\cite{Darlington20,Rodriguez21} via photoluminescence studies. Note that for bubbles with higher aspect ratios the shape is more triangular.\cite{Carmesin19}

We then insert the microscopically obtained 0D potential in the Schr\"odinger equation using 
for electrons in K valley standard single-particle dispersion \cite{Xiao12,Liu13} and scalar plane-waves approximation.\cite{Rosati18}
Besides the (quasi-)continuous states with positive energies we find 5 bound states at the energies $E_1\equiv E_s =-24.4$~meV, $E_2\approx E_3\equiv E_p=-17.1$~meV and $E_4\equiv E_d^{(1)}=-5.81$~meV and $E_5\equiv E_d^{(2)}=-5.77$~meV.
The wave functions of the $s$ and $p$ states are plotted in Figs.~\ref{fig1n}(e-g) and additionally a cross section along the line $x=0~$nm of the $s$ and 
one $p$ state is shown together with the potential in Fig.~\ref{fig1n}(c). One can see a slight tilt of the $p$ states with respect to the $x$ and $y$ axis and 
also a slight asymmetry of $\Psi_3$ with respect to $y$ due to small deviations from radial symmetry [cf. Figs.~\ref{fig1n}(c,g)]. 

These states now act as a basis for the density matrix treatment to describe the dynamics of the electronic density in the 2D-0D TMDC system. 
Mimicking a strongly-localized optical excitation, we construct an electronic wave packet as initial condition. While different amplitudes and 
orientations will be used, in the following all wave packets have the same excess energy as well as energetic and spatial width (see Supporting Information for additional information). The  excess energy of $E_0=22.5$~meV 
corresponds to a velocity of $v \approx  127.4$~nm/ps (for a possible experimental realization see Supporting Information).  

We then set up the equation of motion for the density matrix including the carrier-phonon interaction. In particular, here we account for the electron-phonon interaction with longitudinal optical (LO) 
phonons of energy $\hbar\omega_{LO}=46.3$~meV via the Fr\"ohlich coupling.\cite{Kaasbjerg12,Sohier16} We use a Lindblad formalism \cite{Rosati18} including 
all non-diagonal density matrix elements accounting for the spatial inhomogenity in our system \cite{Rossi02}, capturing most effects found also in 
quantum kinetics calculations  (for more details see Refs. \cite{Reiter06,Rosati17,Rosati18,Lengers20} 
and Supporting Information). We note that for the traveling wave packet we consider coherent dynamics in contrast 
to diffusion dynamics already observed in experiments.\cite{Kulig18,Cadiz18,Perea19,Wang19,Cordovilla19,Zipfel20,Chu20,Rosati21a}
Here in fact we address the dynamics in the very first few hundreds of femtoseconds at very low temperature, where TMDCs show scattering times of several picoseconds \cite{Wagner21} and energy-thermalization timescales of tens of picoseconds, as recently experimentally observed for related excitons \cite{Rosati20b} via phonon-assisted photoluminescence, cf. \cite{Brem19}. The slow intravalley thermalization observed at low temperatures is due to the decreased effectiveness of scattering with phonons, as observed optically via vanishing associated contribution to the excitonic linewidth.\cite{Selig16,Christiansen17,Cadiz17b,Lengers20abs} For these reasons  we  restrict ourselves to the coherent limit for the traveling distribution, while due to energy separations the capture is typically ruled by emission of optical phonons,\cite{Reiter06,Reiter07,Reiter09,Rosati17} as also shown recently for excitons in Mo-based TMDC monolayers.\cite{Lengers20} Since  we discuss the capture of electrons,  we focus on the Fr\"ohlich coupling
which is particularly efficient in these materials.\cite{Sohier16} We stress that we focus on the low-density regime, so that Pauli-blocking or Coulomb-induced effects \cite{Lengers19} can be neglected.

When the wave packets impinge on the nanobubble, carriers are captured into the bound states via the electron-phonon interaction. 
Previous studies have shown that besides energy selection rules, the carrier capture crucially depends on the spatial characteristics, 
as it can happen only when the wave packet is in the vicinity of the nanobubble, i.e. the capture is \textit{local},\cite{Reiter06,Rosati17,Rosati18} 
and it is affected by its direction of motion.\cite{Rosati18} 
More details on the methods and simulation can be found in the Supporting Information and in Refs.~\cite{Carmesin19} and \cite{Rosati18}.

\begin{figure*}[t!]
\centering
 \includegraphics[width=0.90\textwidth]{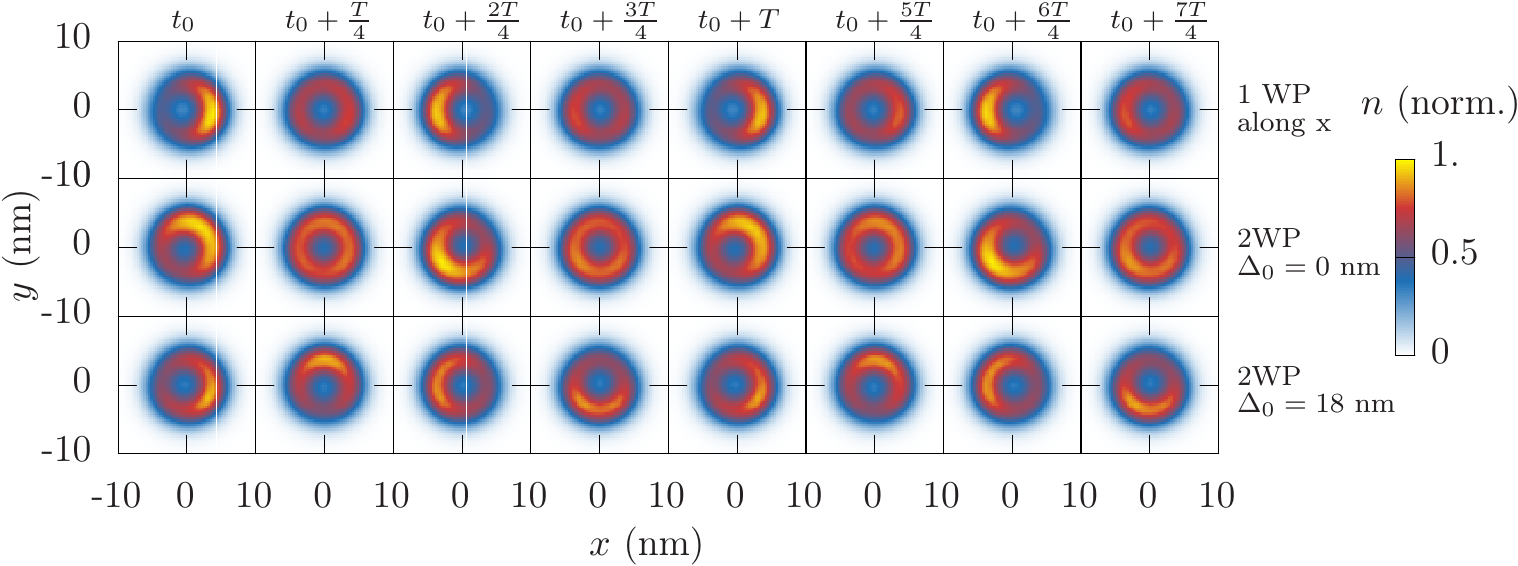}
\caption{Snapshots of the electronic density $n(x,y,t)$ inside the nanobubble for different scenarios.  A single 
wave packet along $x$-direction (top), two wave packets from $x$ and $y$ with $\Delta_0=0$~nm (middle) and two wave packets with $\Delta_0=18$~nm$\equiv \Delta_T/4$ (bottom).}
\label{fig:dyn}
\end{figure*}

\section{Dynamics at the bubble}

During passage of the wave packets, electrons with a density of $n(x,y,t)$ are captured into the nanobubble.\cite{Rosati18}
Figure~\ref{fig:dyn} shows snapshots of the captured density for different scenarios. The time $t_0$ takes into account that 
the wave packet needs a certain time to reach the nanobubble where it can be captured. Using a typical starting distance $x_0=y_0=45$~nm, 
the time is given by $t_0=x_0/v\approx$ 350 fs (cf. also Fig. S2 in Supporting Information) and analogous for $y_0$.

In  Fig.~\ref{fig:dyn}(top row) the effect of  a single wave packet moving along the $x$-direction and crossing the nanobubble
is shown. The captured electronic density displays an oscillatory movement along the $x$-axis. 
The oscillation results from the capture into a superposition state of the localized p-states and the s-state.
Therefore the period agrees well with the value provided by the energy difference between $s$ and $p$ state, i.e. $T = 2\pi \hbar/(E_p-E_s)=0.565$~ps. For a different potential, a similar movement was discussed in Ref.~\cite{Rosati18}.
In the following, we use two wave packets of the same amplitude: 
one traveling along $x$ and one traveling along $y$ (see Fig.~\ref{fig1n}(a)). The dynamics of the density inside the nanobubble is controlled  by adjusting the initial distance of the two wave packets from the nanobubble. We define the difference in starting distances as $\Delta_0=y_0-x_0$. Note that other control scenarios are also possible, e.g., by changing the respective starting time. A spatial separation $\Delta_0$ corresponds 
to a temporal separation $\Delta_0/v$ between the moment in which the corresponding wave packets impinge on the 
nanobubble. Thereby, $\Delta_T=vT\approx 72$~nm marks the distance 
corresponding to the travel of a wave packet during a period $T$.
Figure~\ref{fig:dyn}(middle row) shows the captured density from two wave packets with $x_0=y_0$, i.e., $\Delta_0=0$. 
The density oscillates in a diagonal movement between the upper right and lower left corner.  
Using a finite difference $\Delta_0=y_0-x_0=18$~nm$=\Delta_T/4$ corresponding to a time $T/4$ we can induce a anti-clockwise circular movement, shown in Fig.~\ref{fig:dyn}(bottom row).
We stress that such a movement cannot be induced by a single wave packet alone. 

To quantify these movements, we define the center-of-mass (COM) as 
\begin{eqnarray*}
	\langle\mathbf{r}(t)\rangle\equiv && \left(\langle x(t) \rangle,\langle y(t) \rangle\right) \quad \text{with}  \\
\langle x(t) \rangle && 
	= \frac{\int x \, n d\mathbf{r}}{\int n d\mathbf{r}}
\end{eqnarray*}
and analogously for $\langle y(t)\rangle$. The evolution of the COM $\langle\mathbf{r}(t)\rangle$ for different $\Delta_0$ is shown in 
Fig.~\ref{fig3n}, where the color encodes the increasing time.


Depending on the number of wave packets and their starting distance $\Delta_0$, we obtain different types of 
movement: For a single wave packet along $x$, the COM-movement is just along $x$ as shown in Fig.~\ref{fig3n}(a), while $\langle y(t) \rangle\approx 0$: 
This happens because the initial configuration is symmetric along $y$, and such a symmetry is preserved by the captured density only if the latter has no 
oscillations in $y$ \cite{Rosati18}. Note that the residual slight deviation from strictly null value of $\langle y(t) \rangle$ can be traced back to the 
slight asymmetry of the wave function $\Psi_3$ in Figs.~\ref{fig1n}(c,g).
Obviously, it is possible to induce a vertical oscillation in a completely analogous manner by excitation along the $y$-axis. This is shown in Fig. \ref{fig3n}(b). 

For the cases shown in the middle and lower row of Fig.~\ref{fig:dyn}, two wave packets with $\Delta_0=0$~nm and $\Delta_0=18$~nm, respectively, 
we find a diagonal movement in Fig.~\ref{fig3n}(c) and a circular movement in Fig.~\ref{fig3n}(e). If we now increase the starting difference further, 
for $\Delta_0=36$~nm we find an anti-diagonal movement in Fig.~\ref{fig3n}(d) and for $\Delta_0=54$~nm a left-circular movement in Fig.~\ref{fig3n}(f). 
These distances correspond to time shifts of multiples of $T/4$, namely, $0$, $T/4$, $T/2$ and $3T/4$ for (c), (e), (d) and (f), respectively. 
For separations of $\Delta_0$ that are not a multiple of $T/4$ an elliptical movement is obtained.

\section{Electronic motion as a two-level system}

 \begin{figure*}[t!]
\centering
 \includegraphics[width=0.80\textwidth]{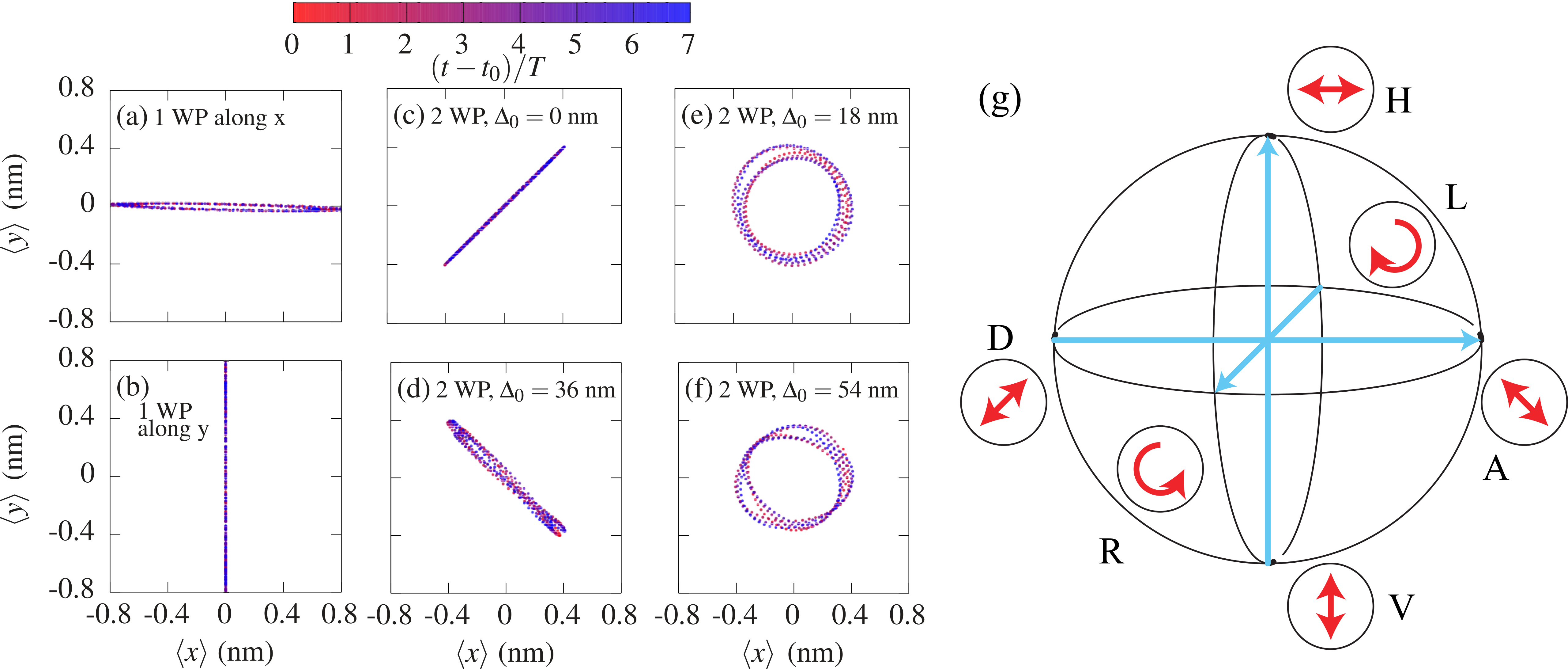}
\caption{(a-f) Dynamics of the COM for the density inside the nanobubble. Left column: a single wave 
packet (1 WP) impinging from the $x$ and $y$-direction. Middle column: two wave packets (2 WP) for 
distances $\Delta_0=0$~nm and $\Delta_0=36$~nm$\equiv \Delta_T/2$. Right column: $\Delta_0=18$~nm$\equiv \Delta_T/4$ 
and $\Delta_0=54$~nm$\equiv 3 \Delta_T/4$. The time is encoded in the color which changes from red (early times) 
to blue (later times). (g) Electronic Poincar\'e sphere with the distinguished states $H$ and $V$ as 
well as the superposition states $D$, $A$, $R$ and $L$ \label{fig3n}}
\end{figure*}

Modern quantum technologies rely on  properly-defined Poincar\'e spheres \cite{} e.g. as defined via 
the different 
light polarizations.\cite{Born13,Verma}  As sketched  in Fig. \ref{fig3n}(g), a Poincar\'e sphere displays horizontal and 
vertical polarizations on the poles ($\ketv{H}$ and $\ketv{V}$) while showing on the equator diagonal/anti-diagonal ($\ketv{D}$ and $\ketv{A}$) as well as left- and right-circular polarizations ($\ketv{L}$ and $\ketv{R}$). Remarkably, these points resemble the electronic COM 
found in Fig.~\ref{fig3n} (cf. red arrows in Fig. \ref{fig3n}(g) with Figs. \ref{fig3n}(a-f)). Starting with these empirical considerations, 
we exploit the above-discussed density dynamics to introduce an \emph{electronic Poincar\'e sphere} analogous to the one defined e.g. by light polarization.

Considering that only the $s$ and two $p$ states contribute to the dynamics within the nanobubble, 
the spatial density $n(\mathbf{r},t)$ in the nanobubble can be separated into three parts
\begin{equation}\label{n}
n(\mathbf{r},t)=n_S(\mathbf{r},t) +n_H(\mathbf{r},t)+n_V(\mathbf{r},t) \quad ,
\end{equation}
where $n_S$ is the contribution not displaying any dynamics after the capture is completed. $n_H$ and $n_V$ are the 
electronic motions in the horizontal and vertical direction, which can be written as
 \begin{equation}\label{n_SHV}
\begin{split}
n_H(\mathbf{r},t) =&c_H(\mathbf{r}) \cos[\omega(t-t_H)] , \\
n_V(\mathbf{r},t) =&c_V(\mathbf{r}) \cos[\omega(t-t_V)] \quad ,
\end{split}
\end{equation}
with the spatial profiles $c_{H,V}$ and time shift $t_{H,V}$ and the frequency $ \omega =2\pi/T$ (see Supporting Information for more details). 
Note that we can map the spatial profiles by using the occupations and coherences of the eigenstates in the nanobubble potentials as outlined in Supporting Information. 

If we consider two equally shaped wave packets impinging from $x$ and $y$ direction, the central factor for the electronic motion within the nanobubble 
are the times when they impinge on the nanobubble. 
The time difference $t_H-t_V$ introduces a phase difference between the horizontal and vertical oscillation according to Eq. (\ref{n_SHV}), which determines the 
movement achieved in the nanobubble.

From Eqs.~\eqref{n} and \eqref{n_SHV} we can extract the COM-movement to
\begin{eqnarray}\label{COM_osc}
	\langle x(t) \rangle =&& X \cos[\omega(t-t_H)]\ ,  \\
	\langle y(t) \rangle =&& Y \cos[\omega(t-t_V)]\ , \notag
\end{eqnarray}
%
%
normalized to the total captured density 
\[ \bar{n} = \int n(\mathbf{r},t) d^2r \,.
\] 
We here used that $\int x c_V(\mathbf{r})d^2r\approx \int y c_H(\mathbf{r})d^2r\approx 0$.

The two-dimensional vector $\langle\mathbf{r}\rangle$ in Eq. (\ref{COM_osc}) can be written as
\[
 \langle\mathbf{r}\rangle\equiv\Re[\mathbf{R} \,\exp[-i \omega t]]
 \]
with $\mathbf{R}=(Xe^{i\omega t_H},Ye^{i\omega t_V})$ and $\Re$ denoting the real part. The vector $\mathbf{R}$ is an electronic analog of the Jones vector,\cite{Verma} 
which is one of the possible ways for describing the light polarization. Another option to describe the polarization of light is the Stokes vector.\cite{Born13,Verma} 
Also for the Stokes vector we can introduce the electronic spatial analog using
\begin{equation}\label{Stokes}
\begin{split}
s_0  =& X^2+Y^2 \\
s_1  =& X^2-Y^2 \\
s_2  =& 2 XY \cos[\omega(t_V-t_H)] \\
s_3  =& 2 XY \sin[\omega(t_V-t_H)] \,. \\
\end{split}
\end{equation} 
From the four Stokes parameters one can define the electronic Poincar\'e sphere with squared radius $s^2_0=s^2_1+s^2_2+s^2_3$ as sketched 
in Fig. \ref{fig3n}(g). From the electronic Poincar\'e sphere we can define the generic state  $\vert \psi \rangle$ of the electronic oscillation as \cite{Verma}
\begin{equation}\label{ket}
\vert \psi \rangle= e^{ i \chi} \left( \cos\left[\frac{\theta}{2}\right] \vert H \rangle + e^{i \phi}\sin\left[\frac{\theta}{2}\right] \vert V \rangle \right) \quad ,
\end{equation}
with 
\begin{equation}\label{ket_par}
\theta=\text{arctg}\left(\frac{Y}{X}\right) \, , \, \chi=-\omega t_H \, , \, \phi=\omega(t_V-t_H) \, ,
\end{equation}
where $0\leq\theta\leq\pi$. Equation (\ref{ket}) is a formal way to associate a quantum-mechanical 
state on the electronic Poincar\'e sphere with the real oscillations of the captured electronic density. 
The density oscillations can thereby be unambiguously mapped onto a Poincar\'e sphere considering the amplitudes $X$ and $Y$ and the phase difference $\omega(t_V-t_H)$.

The Jones vector, Stokes coefficients and electronic Poincar\'e sphere are all analogous ways to describe a two-level system of the electronic spatial 
oscillations inside the nanobubble, with e.g. $\vert H \rangle$ being the ground state and $\vert V \rangle$ the excited one. 
Of course different choices of basis states are possible.\cite{Verma} We stress that the electronic state in Eq.~\eqref{ket} per se is time independent, 
as the Jones vector, Stokes coefficient and Poincar\'e-sphere points do not depend on time $t$, but only depend on time shifts $t_H$ and $t_V$. 
Nevertheless, its manifestation are time-dependent oscillations of $n(\mathbf{r},t)$ or, analogously, of $\langle \mathbf{r}\rangle$. 
Such an electronic two-level system could be useful e.g. to store information in a stationary qubit. 

\section{State switching}
\label{sec:control}

\begin{figure}[t!]
\centering
 \includegraphics[width=\columnwidth]{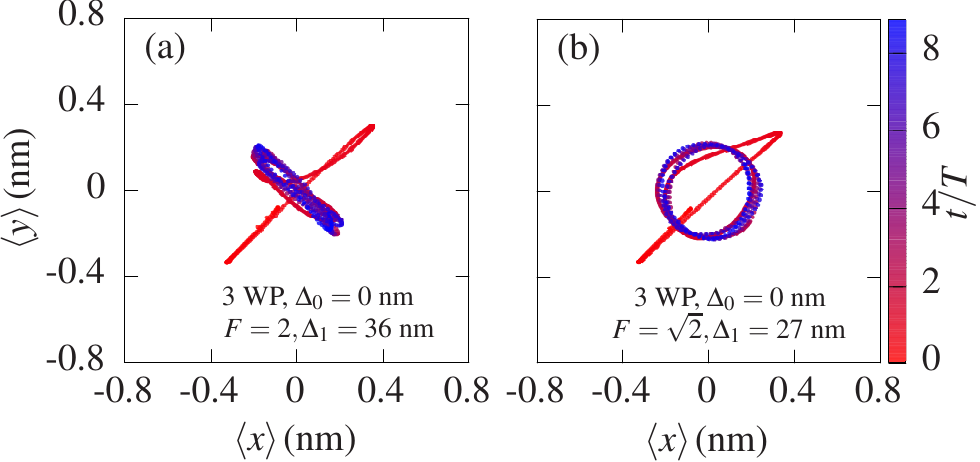}
\caption{Dynamics of the COM for the electronic density captured inside the TMDC 
nanobubble induced by three wave packets, where the first two start equally with $\Delta_0=0$~nm, 
while the third one is by a factor of $F$ bigger and comes from the $y$-direction with $\Delta_1=36$~nm (left) and  $\Delta_1=27$~nm (right). 
The time is encoded in the color which changes from red (early times) to blue (later times).  \label{fig:control}}
\end{figure}

The proposed mechanisms can go beyond the \textit{initialization} of the electronic Poincar\'e sphere by \textit{switching} its state. To show this we start initializing a 
diagonal movement with $\Delta_0=0$. 
Then we add a third wave packet traveling along the $y$-direction, which is shifted w.r.t. the first one by $\Delta_1=x_0-y_1$ and has an amplitude bigger by a factor $F$  than the amplitude of the first two. 

The  dynamics of the COM for the electronic density captured inside the TMDC nanobubble is shown in Fig.~\ref{fig:control}. 
In both cases it starts with the diagonal movement induced by the first two wave packets (red dots  between first and third quadrant in Figs. \ref{fig:control}(a,b)). 
In the first control example of Fig.~\ref{fig:control}(a) the third wave packet has twice the intensity of the first two  ($F=2$) and starts with $\Delta_1=vT/2=36$~nm. As a result, the initial diagonal movement is switched 
into the anti-diagonal one (blue dots between second and fourth quadrant). The simulation results are consistent with previous considerations: 
While the first two wave packets induce a state $e^{i \chi}(\ketv{H}+\ketv{V})$, the third one induces an additional 
contribution $Fe^{i \chi}(\exp[-i\omega T/2]\ketv{V})\equiv-2e^{i \chi}\ketv{V}$. 
Note that we do not use normalized states here because in the end the state must be normalized to the total density 
present in the system (thereby defining the radius of the Poincar\'e sphere). 
The first two wave packets are initialized with $F=1$ in $x$ and $y$ direction and the second one with $F=2$ in $y$-direction, 
such that the sum thus provides $e^{i \chi}(\ketv{H}-\ketv{V})\equiv \ketv{A}$.  We then have 
\begin{equation*}
	 |D\> \stackrel{\frac{1}{2}n_V(T/2)}{\longrightarrow} |H\> \stackrel{\frac{1}{2}n_V(T/2)}{\longrightarrow} |A\> \quad .
\end{equation*}
Note that the overall amplitude of the COM has decreased by approximately a factor of 2 w.r.t. to 
the initial state $\ketv{D}$ due to the normalization mentioned above, cf. Eq.~\eqref{COM_osc} and Fig. \ref{fig:control}(a).

In the second example we set $F=\sqrt{2}$ and $\Delta_1=3vT/8=27~$nm, corresponding to an additional 
phase $\bar{\varphi}=-3\pi/4$ of the third wave packet w.r.t. the first two. The result is 
shown in Fig.~\ref{fig:control}(b), which shows a right-circular motion. The $R$ state is expected,
because besides state $|H\> $ we can calculate in the vertical direction $(1+Fe^{- i \bar{\varphi}})|V\>\equiv -i|V\>$. 
Even more versatile controls can be obtained by adding additional wave packets.
This shows that the wave packets can fully control the electronic Poincar\'e sphere via both initialization and switch,
indicating the versatility of the proposed two-level system.

\section{Conclusion}

In conclusion, we have demonstrated  an electronic  Poincar\'e sphere, whose states can be initialized and switched by capture from traveling wave packets.  For this purpose we combine the first-principle description of a TMDC nanobubble and associated confining potential with a density-matrix description of ultrafast electronic transport and capture.
Thanks to the locality of the carrier capture
the non-trivial dynamics of the captured distribution can be fully controlled. 
We find that
the phase difference between horizontal and vertical oscillation is analogous to the polarization states of light, i.e.  
the Poincar\'e sphere, and fully controllable by the time delay of additional wave packets.
  Our results show how new opportunities of information protocols could be offered by the miniaturization properties of two-dimensional materials 
and the spatial control of carrier dynamics.

\section*{Acknowledgements}
We thank F. A. Reiter for graphical help with Fig.~1. 
R.~R. acknowledges funding from the Deutsche Forschungsgemeinschaft
(DFG) through SFB 1083 (subproject B9) and
the European
Union's Horizon 2020 research and innovation programme
under grant agreement no. 881603 (Graphene Flagship). 
F.~L. and D.~E.~R. acknowledge financial support by the Deutsche Forschungsgemeinschaft (DFG) 
through the project 406251889 (RE 4183/2-1). C.~C., M.~F., F.~J., and M.~L. acknowledge 
 support by the Deutsche Forschungsgemeinschaft (DFG) within
RTG 2247 and through a grant for CPU time at the HLRN (G\"ottingen/Berlin).

\appendix

\section{Simulation details} \label{sec:simulation}
\label{app:simulation}

Our simulation is a multi-step process: (i) We perform a microscopic calculation of the nanobubble potential
based on a combination of valence force field and 6-band tight binding
model. (ii) The nanobubble potential is inserted into the Schr\"odinger equation
for K-valley electrons, yielding bound and free states together with the
corresponding energies. (iii) Using the wave functions obtained from the Schr\"odinger equation as
basis states, we set up the equation of motion for the electron density
matrix in the framework of a Lindblad superoperator technique. (iv)  As initial condition we construct wave packets in the subspace of
the free states localized at a certain distance from the nanobubble and
calculate their time evolution according to the density matrix equation. We here describe these steps in detail:

Regarding step (i): For the simulation of the nanobubble we use a combination of atomic force field calculations with tight binding simulations. 
Starting from a paraboloid profile of the nanobubble, the atoms in the upper TMDC layer are relaxed, while the atoms in the lower layer are kept fixed. 
This part of the simulation is performed using a REAX potential \cite{vanDuin01} with the parametrization from Ref.~\cite{Ostadhossein17} within a 
valence force field calculation, which is capable of accurately describing bond deformations under strain as well as continuous bond formation and breaking dynamics. 
Based on these valence force field simulations, new equilibrium positions of the individual atoms in the bended material are determined. 
Information about the displaced atoms is used in a subsequent tight-binding electronic-state calculation for the supercell structure. 
For a supercell with an in-plane extension of 130$\,$nm and up to 1.2$\times10^6$ atoms, a 6-band tight-binding Hamiltonian is solved as discussed 
in Ref.~\cite{Carmesin19}. Strain-induced local band gap changes arising from the displaced atomic positions are included via a generalized 
Harrison rule~\cite{Froyen79}. 

Additionally, when locally detaching the upper layer from the substrate underneath and changing from a commensurate bilayer to monolayer-like structures 
across the nanobubble, the upper layer is subject to a modified dielectric environment and electronic hybridization 
is hence different than in a bilayer~\cite{Cheiwchanchamnangij12, Roesner16, Raja17, Florian18}. 
As discussed in Ref.~\cite{Carmesin19}, these effects are included in the calculation by changing individual tight-binding parameters.

Regarding step (ii): We insert the potential evaluated in step (i) in a Schr\"odinger equation where we assume that the eigenstates for the bare-monolayer behave around the K valley as scalar plane waves with free eigenenergies $E_{\mathbf{k}}=(\lambda_c+\lambda_v)/2+a_0 t \sqrt{[(\Delta_G-(\lambda_v-\lambda_c))/(2a_0t)]^2 + |\mathbf{k}|^2}$ \cite{Rosati18}, with $a_0=3.193$ \AA, $\Delta_G=$1.66 eV, $t=1.1$ eV, $\lambda_v$=75 meV and $\lambda_c=-1.5$ meV \cite{Xiao12,Liu13}, cf. also Ref. \cite{Rosati18}.

Regarding step (iii): 
We further set up the equations of motion taking into account the free evolution of the electronic wave packet (see point (iv) below for its initial definition) and scattering with LO phonon via Fr\"ohlich interaction. 
While for excitons intervalley scattering can be very important \cite{Selig16,Brem19,Rosati20b,Rosati21a}, despite efficient Hamiltonian coefficients \cite{Jin14} free electrons close to the K-valley energy mininum cannot directly scatter in the free states of other valleys due to energy conservation or uneffectiveness of absorption of phonons with finite energies at cryogenic temperatures. In view of the  picoseconds scattering times \cite{Wagner21} and tens of picoseconds energy-thermalization time of related excitonic states \cite{Rosati20b}, intravalley scattering with acoustic phonon of K electrons is expected to take place at longer timescales than the ones here considered, hence it is neglected. As a consequence we restrict ourselves to the free coherent evolution for the free distribution. The scattering between free and bound states is dominated by optical phonon modes due to energy separation between emitting and receiving states\cite{Reiter06,Reiter07,Rosati17}, as shown recently for excitons in Mo-based TMDC monolayers \cite{Lengers20}. In particular here we focus on the effective scattering with LO phonon via Fr\"ohlich interaction, whose 
 scattering coefficients between electronic 
states in valley K with momenta $\mathbf{k},\mathbf{k+q}$ are given by $g_{\mathbf{q}}\equiv \langle \mathbf{k+q} \vert \hat H_{e-LO} \vert \mathbf{k}\rangle$, where the electron-LO phonon Hamiltonian $H_{e-LO}$ follows Ref.~\cite{Sohier16}. 
These scattering coefficients have been inserted in a  Lindblad approach for the carrier capture \cite{Rosati17,Rosati18}, which is able to describe the 
scattering from the delocalized states into the localized ones not only tracking the energy of the emitting states, but also the spatial information. 
Key to describe the dynamics in this highly inhomogeneous case is the off-diagonal nature of the Lindblad superoperator together with an energetic broadening of 
the energy conservation, the latter previously included  by comparison with the full quantum kinetics calculations \cite{Rosati17}, 
although more recent studies have shown similar broadenings \cite{Lengers20}.

\begin{figure}
\includegraphics[width=0.9\columnwidth]{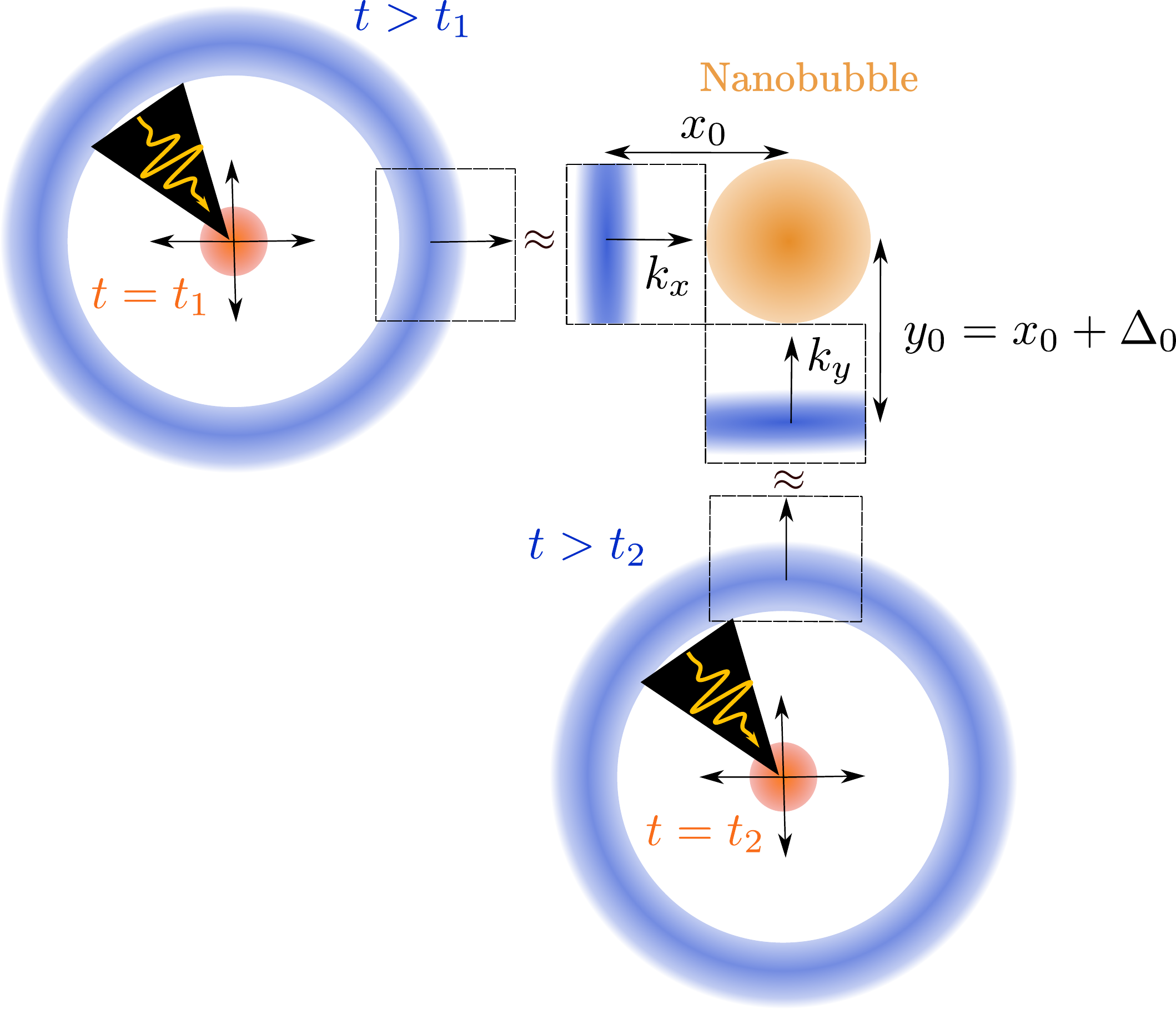}
\caption{Possible experimental realization of the considered setup. Two strongly localized excitation sources lead to the considered 
wave fronts moving in $x$ and $y$ direction (marked parts of the ring-like wave packets). A temporal delay $t_2-t_1$ between the excitation of the two 
wave packets can be used to induce a difference in starting positions $\Delta_0=y_0-x_0$ as used in the simulations.}
\label{fig:app_exp}
\vspace{-0.3cm}
\end{figure}

Finally, regarding point (iv) we choose a wave-front type packet, which in the density matrix and in the basis of the free TMDC states and for the $x$-propagating 
case can be written as \cite{Rosati18}
\begin{equation*}
\begin{split}
    \rho_{\mathbf{k} + \frac{\mathbf{k}'}{2},\mathbf{k} - \frac{\mathbf{k}'}{2}} \propto & e^{-\frac{1}{2}(k'_x \Delta)^2} e^{- i  k'_x x_0}\times \\
     & e^{-\left(\frac{(\hbar^2
k_x^2)/(2 m^*) -E_0}{\sqrt{2}\Delta_E}\right)^2}  \Theta(k_x) \delta(k'_{y})
\end{split}
\end{equation*}
where $\mathbf{k}\equiv(k_x,k_y)$ is the 2D wave vector and $\Theta(k_x)$ is the Heaviside step function. 
The wave packet propagating along $y$ has the same form with $k^\prime_x \leftrightarrow k^\prime_y$ and ($x_0 \leftrightarrow y_0$). The wave packets have a finite width in space and energy of $\Delta$= 10 nm and $\Delta_E$ = 5 meV, respectively, and are centered at $\mathbf{r}_0\equiv (x_0,0)$ or ($0,y_0$) for propagation along $x$ and $y$, respectively. 
The excess energy, which determines the velocity as well as the bound states which are energetically favorable for the carrier capture, 
is taken to be $E_0 = 22.5$ meV, i.e., an energy close to $(E_s+E_p)/2+$E$_{LO}$, with $E_s$ and $E_p$ being the energy of state 1 and nearly degenerate states 2 and 3, respectively (see main manuscript). Such a wave packet can in principle be found in certain distances from the 
excitation spot of a strongly localized near-field source \cite{Rosati18}. The energetic width $\Delta_E=5~$meV corresponds then to an ultrafast 
excitation of about 150 fs duration. These length and timescales could be combined as typically done in the field of ultrafast nano-optics (see, e.g., \cite{Vasa09} for 
a review on experimental applications). These experimental developments have led to extensive studies of nanometric wave packets (see, e.g., \cite{Reichelt09,Rosati13b,Rosati15}). 
A specific setup for the presented system is given in Fig.~S\ref{fig:app_exp}, where two near-field sources lead to the considered wave packets in $x$ and $y$ direction 
impinging on the nanobubble. In this setup a temporal delay $t_2-t_1$ between the exciting pulses can be used to effectively tune the difference in starting 
positions $\Delta_0$.

Note that we work in the single-band approximation with the standard single-particle dispersion \cite{Xiao12,Liu13}, while also excitonic effects in 
TMDC monolayers are very prominent \cite{Wang18,Mueller18}. 
Nevertheless beside the formation  of deeply-bound excitons \cite{Selig18,Brem18}, in competition with spatial separation between electrons and holes \cite{Lengers19}, we also expect
electronic wave packets able to travel through the monolayer and then be captured in the here-considered ultrafast timescale.

\section{Occupations and coherences of the captured electrons}\label{app:theory}
While in the manuscript we focus purely on the description using the real-valued density, it is interesting to consider the occupation and the 
coherences of the bound states. More details on the quantum mechanical description of the states using the density matrix formalism can be found 
in previous publications \cite{Rossi02,Reiter06,Reiter07,Rosati14,Rosati17,Rosati18}.

In the density matrix formalism the single particle density matrix within the eigenbasis states $|i\rangle$ is given by
\[\hat \rho\equiv \sum_{i,j} \rho_{ij}(t)\vert i \rangle \langle j \vert
\] 
from which the spatial distribution is calculated via
\begin{equation*}
    \begin{split}
        n(\mathbf{r},t) \equiv &\sum_{i,j} n_{ij}(\mathbf{r},t)\quad \text{with }  \\
        n_{ij}(\mathbf{r},t) = &\rho_{ij}(t)\Psi_i(\mathbf{r}) \Psi^*_j(\mathbf{r}).
    \end{split}
\end{equation*}
  
In our case the states $i=1,\ldots 5$ correspond to the bound states in the nanobubble potential (see discussion in the main text). Since for our initial conditions the the occupation of the states $\Psi_4$ and $\Psi_5$ is negligible, we will restrict our discussion to states $i=1,2,3$.

The occupations and coherences of the states are then given respectively by
\[
 f_i = \rho_{ii}=\langle i \vert \hat \rho \vert i \rangle \quad \text{and} \quad p_{ij}=\rho_{ij,i\neq j}=\langle i \vert \hat \rho \vert j \rangle _{i\neq j} \quad .
\]
The most remarkable feature is that the capture from the 2D system into the 0D localized potential results not only in occupations, but in a superposition state of the different states \cite{Rosati18}. The general density matrix reads
\begin{equation}\label{DM}
\hat \rho =\frac{1}{2} \sum_{i=1}^3\! f_i \vert i \rangle \langle i \vert + e^{- i \omega t} \sum_{i=2}^3 \tilde{p}_{1i} \vert 1 \rangle \langle i \vert + \text{H.c. ,}
	\end{equation}
with $\omega=(E_s-E_p)/\hbar$ and H.c. denoting the Hermitian conjugate, and where we introduced $p_{1i}=\tilde{p}_{1i}\exp\left(-i\omega t\right)$ in the interaction picture. This equation already shows that the capture into the superposition state results in a dynamics of the captured density with the frequency $\omega$. The shape of the dynamics, however, is determined by the magnitude of the occupations $f_i$ together with the absolute values of the coherences $p_{12}$ and $p_{13}$ and most importantly the relative time difference between the coherences, which in turn is determined by the time of the carrier capture.  

\begin{figure}[h!]
\centering
 \includegraphics[width=0.9\columnwidth]{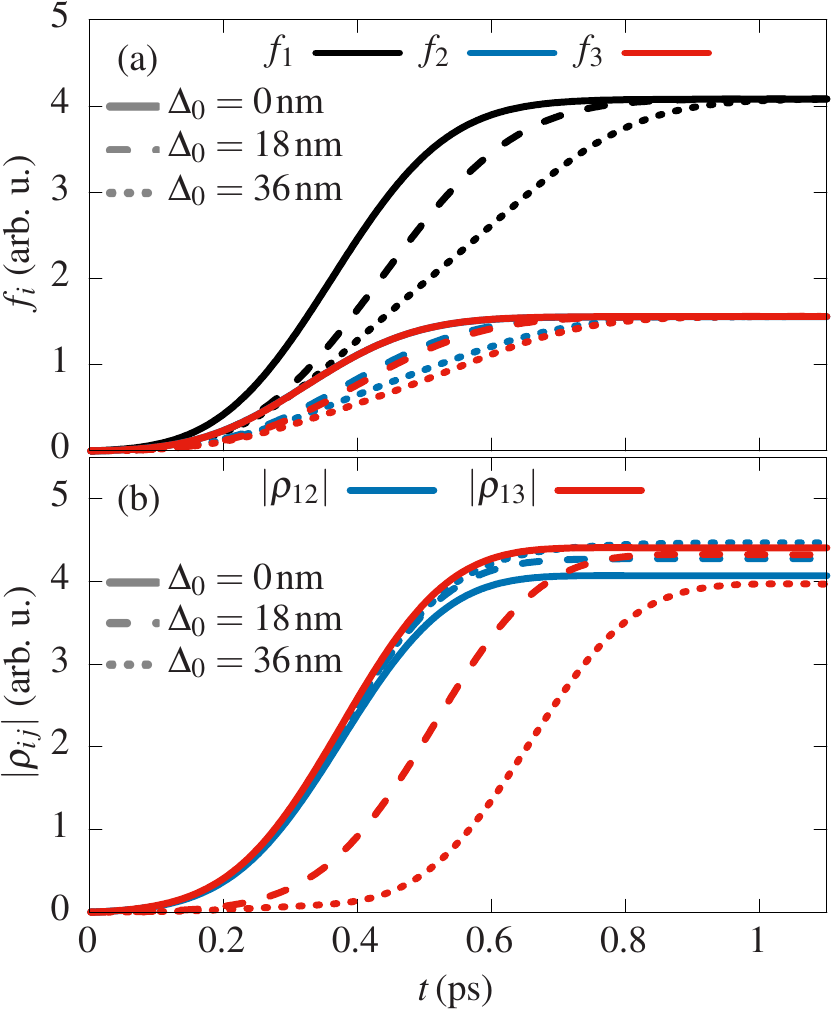} 
\caption{ Dynamics of (a) the occupations of the bound states and (b) the absolute values between the coherences of the bound 
states for excitation with two wave packets from $x$ (initial distance from the nanobubble $x_0$=45 nm) and $y$-direction with different starting distances $\Delta_0=y_0-x_0=0,18$ and $36$ nm. 
These correspond to the dynamics shown in Fig.~3(middle, bottom row) and  COM movements seen in Fig.~4(c), (e) and (d) of the main manuscript, 
respectively.\label{fig:appendix}}
\end{figure}

One example of carrier capture dynamics is shown in Fig.~S\ref{fig:appendix}, where the dynamics of (a) the occupations and (b) the coherences is shown. 
We stress again that the capture happens locally, i.e., when the wave packet is close to the nanobubble as can be seen in the corresponding 
finite rise time of the occupations \cite{Rosati17}. The final capture is slightly stronger into state $1$ then into the 
excited states $2$ and $3$ as seen in Fig.~S\ref{fig:appendix}(a), while the coherences after the capture in Fig.~S\ref{fig:appendix}(b) are 
equally strong. However, the build-up of the population and in particular the coherences is different: The wave packet traveling along the $x$-direction 
induces the coherence $p_{12}$, while the wave packet traveling along $y$-direction induces $p_{13}$. When the latter wave packet is delayed, $p_{13}$ builds up later, 
but the final value $|p_{13}(t\rightarrow\infty)|$ is almost independent of $\Delta_0$. One may therefore approximate 
the dynamics of $p_{13}$ as $p_{13}\approx |p_{13}(t\rightarrow\infty)|\Theta(t-t_V)\exp\left(-i\omega(t-t_V)\right)$ with the 
Heaviside function $\Theta$ when approximating the build-up of coherence as instantaneous at the arrival time $t_V=y_0/v$ of the 
wavepacket in $y$-direction. Accordingly we find for these three examples the diagonal movement, the circular movement and the anti-diagonal movement 
due to the phase difference $\omega(t_H-t_V)=\omega\Delta_0/v$.

From the microscopic picture given in Eq.~(\ref{DM}) we obtain the spatial profiles of the oscillation of the density as defined in Eq. (2) of the main manuscript. 
In particular we find that $c_{V}=2\left|p_{13}(t\rightarrow\infty)\right|\psi_1(\mathbf{r})\psi_{3}(\mathbf{r})$ and $t_{V}=y_0/v$ with the starting 
distance $y_0$ from the QD and replacing $3 \to 2 $ we get similar expressions for $c_H$ and $t_H$. This connects the quantum mechanical picture of 
the states on the Poincar\'e sphere to the dynamics of the captured density.


%

\end{document}